\documentclass[a4paper]{article}
\usepackage{amscd,amsmath,amssymb}
\usepackage{yhmath}

\newcommand{\cH}{{\cal H}}

\newcommand{\cQ}{{\cal Q}}
\newcommand{\cbQ}{\overline{\cal Q}}

\newcommand{\cV}{{\cal V}}

\newcommand{\bQ}{{\overline Q}{}}

\newcommand{\bxi}{{\bar\xi}}

\newcommand{\brho}{{\bar\rho}{}}

\newcommand{\cN}{{ {\cal N}   }}

\newcommand{\tPi}{{\widetilde \Pi}}

\def\im{{\rm i}}

\newcommand{\be}{\begin{equation}}
\newcommand{\ee}{\end{equation}}
\newcommand{\bea}{\begin{eqnarray}}
\newcommand{\eea}{\end{eqnarray}}
\newcommand{\ba}{\begin{array}} \newcommand{\ea}{\end{array}}

\newcommand{\und}{\qquad\textrm{and}\qquad}

\newcommand{\p}[1]{(\ref{#1})}

\begin{document}
\thispagestyle{empty}

\begin{flushright}
\end{flushright}
\vspace{2cm}

\begin{center}
{\huge\bf $\cN{=}\,4$ supersymmetric \\[8pt] Calogero--Sutherland models}
\end{center}
\vspace{1cm}

\begin{center}
{\LARGE\bf  Sergey Krivonos${}^{a}$, Olaf Lechtenfeld$^b$}
\end{center}

\vspace{0.2cm}

\begin{center}
{${}^a$ \it
Bogoliubov  Laboratory of Theoretical Physics,\\
Joint Institute for Nuclear Research,
141980 Dubna, Russia}\vspace{0.3cm}

${}^b$ {\it
Institut f\"ur Theoretische Physik and Riemann Center for Geometry and Physics, \\
Leibniz Universit\"at Hannover,
Appelstrasse 2, 30167 Hannover, Germany}
\vspace{0.5cm}

{\tt krivonos@theor.jinr.ru, lechtenf@itp.uni-hannover.de}
\end{center}
\vspace{3cm}

\begin{abstract}
\noindent  
Starting from the Hamiltonian formulation of supersymmetric Calogero models associated with 
the classical $A_n, B_n, C_n$ and $D_n$ series we construct the $\cN{=}\,2$ and $\cN{=}\,4$ 
supersymmetric extensions of the their hyperbolic/trigonometric Calogero--Sutherland cousins. 
The bosonic core of these models are the standard  Calogero--Sutherland hyperbolic/trigonometric systems. 
\end{abstract}


\newpage
\setcounter{page}{1}
\setcounter{equation}{0}
\section{Introduction}
There is a lot of confirmation that $\cN{=}\,4$ supersymmetric extensions of Calogero--Moser systems must include a large number of fermions -- far more than the $4n$  fermions expected within the standard (but not very successful) approach \cite{SF1,SF4,SF5,KLS1,KLS2,KLPS}. The source of these fermions is the supersymmetrization of the matrix
models from which, in the purely bosonic cases, the Calogero--Moser systems can be obtained by a reduction (see e.g.~\cite{Poly2}). 

A suitable approach to supersymmetric Calogero-like models has been proposed in  \cite{SF1,SF4,SF5}. Starting from a supersymmetrization of the Hermitian matrix model, the resulting matrix fermionic degrees of freedom are packaged in $\cN{=}\,4$ superfields.
In a recent paper \cite{SF4}, $\cN{=}\,2$ and $\cN{=}\,4$ supersymmetric extensions of the multiparticle hyperbolic Calogero--Sutherland system were constructed by applying a gauging procedure~\cite{FD} to one-dimensional matrix superfield systems. 
However, for $\cN{=}\,4$ their bosonic part does not reproduce the ordinary Calogero systems but only spin-Calogero ones.\footnote{For a quantization of the $\cN{=}\,2$ case, see~\cite{SF5}.}

In a series of papers \cite{KLS1,KLS2,KLPS} we developed a different approach. Mainly working in the Hamiltonian
formulation, we worked out an ansatz for the supercharges which accommodates all Calogero models associated
with the classical $A_n, B_n, C_n$ and $D_n$ Lie algebras. Here, the supercharges contain the fermion-cube terms only through the combination {\sl fermion $\times$ fermion bilinear}, where the fermion bilinears span 
an $s(u(n)\oplus u(n))$ algebra.

In this paper we use this ansatz to construct $\cN{=}\,4$ supersymmetric extensions of the Calogero--Sutherland models associated with the classical $A_n, B_n, C_n$ and $D_n$ series (Section 2).   As a separate application, we also find the $\cN{=}\,4$ supersymmetric extensions of the trigonometric/hyperbolic cousins of the Euler--Calogero--Moser system (Section 3).

\setcounter{equation}{0}
\section{Calogero--Sutherland  models}
\subsection{Basic ingredients}
The starting point of our  construction is the same set of the fields as in  the $\cN$-extended supersymmetric
Calogero--Moser model~\cite{KLS1} which is nothing but  a supersymmetric extension of the
Hermitian matrix model \cite{sCal, GH, Poly2}. This set of fields includes the following ones
\begin{itemize}
\item $n$ bosonic coordinates $x_i$, which come from the diagonal elements
of the Hermitian matrix $X$, and the corresponding momenta $p_i$ for $i,j=1,\ldots,n$ which obey the standard brackets
\be\label{PBb}
\left\{ x_i, p_j\right\} = \delta_{ij} ,
\ee
\item fermionic matrices containing $\cN\,n^2$ elements $\xi^a_{ij}, \bxi_{ij\,a}$
for $a=1,\ldots,\cN/2$ with $(\xi^a_{ij})^\dagger=\bxi_{ji\, a}$ and brackets
\be\label{PB0f}
\left\{ \xi^a_{ij} , \bxi_{km\, b}\right\} = -\im \, \delta^a_b \delta_{im}\delta_{jk}.
\ee
\end{itemize}

Using these ingredients one may construct the fermionic bilinears
\bea\label{Pi0}
\Pi_{ij} & =&  \sum_{a=1}^{\cN/2} \;\sum_{k=1}^n \left( \xi^a_{ik}\bxi_{kj\,a}+\bxi_{ik\,a}\xi^a_{kj}\right), \quad \sum_i \Pi_{ii} =0,  \\
\tPi_{ij} & = & \sum_{a=1}^{\cN/2} \;\sum_{k=1}^{n } \left( \xi^a_{ik}\bxi_{kj\,a}-\bxi_{ik\,a}\xi^a_{kj}\right),
\eea
which form  an $s( u(n)\oplus u(n))$ algebra,
\be\label{susus}
\bigl\{ \Pi_{ij}, \Pi_{km} \bigr\}=\bigl\{ \tPi_{ij}, \tPi_{km} \bigr\}=
\im \bigl( \delta_{im} \Pi_{kj}-\delta_{kj}\Pi_{im}\bigr) \und
\bigl\{ \Pi_{ij}, \tPi_{km} \bigr\}=\im \bigl( \delta_{im} \tPi_{kj}-\delta_{kj}\tPi_{im}\bigr) \ .
\ee

Using these ingredients in \cite{KLPS} the supercharges and Hamiltonian have been constructed for arbitrary even-$\cN$ supersymmetric extensions
of the $A_n$,  $B_n$, $C_n$ and $D_n$ rational Calogero models. In what follows we will use the same ingredients to construct $\cN=2,4$  trigonometric/hyperbolic Calogero--Sutherland models with the supercharges and the Hamiltonian obeying the $\cN=2,4$-extended super-Poincar\'{e} algebra.

\subsection{$\cN{=}\,2$ supersymmetric  $A_{n-1}\oplus A_1$ Calogero--Sutherland  models}
In this simplest case the supercharges have a quite simple structure\footnote{We omit the indices $a,b$ which all are equal to one in this case.}
\be\label{Q2}
\begin{aligned}
 & Q = \sum_{i=1}^n p_i \xi_{ii}\ -\ \im \sum_{i \neq j}^n \Bigl[ f(z_{ij})\left(g + \Pi_{jj}\right) +\frac{f'(z_{ij})}{f(z_{ij}) }\Pi_{ij}\Bigr]\, \xi_{ji} \ ,& \\
& \bQ  =  \sum_{i=1}^n p_i \bxi_{ii}\ -\ \ \im \sum_{i \neq j}^n \Bigl[ f(z_{ij})\left(g + \Pi_{jj}\right) +\frac{f'(z_{ij})}{f(z_{ij}) }\Pi_{ij}\Bigr]\, \bxi_{ji}  .&
 \end{aligned}
\ee
Note that $\tPi_{ij}$ does not appear here and the function $f$ will be specified in a moment.

These supercharges form an $\cN{=}\,2$ super-Poincar\'{e} algebra 
\be\label{N2SP}
\begin{aligned}
\left\{ Q , \bQ \right\} = - 2 \im\, H \und
\left\{ Q, Q \right\}=\left\{ \bQ, \bQ \right\}=0 \ .
\end{aligned}
\ee
together with the Hamiltonian
\be\label{H2}\
\begin{aligned}
H \ = \frac{1}{2}\sum_{i=1}^n p_i^2\ +\ \frac{1}{2}\sum_{i \neq j}^n \Big[ \left(g+\Pi_{jj}\right) f(z_{ij}) +\frac{f'(z_{ij})}{f(z_{ij})}\Pi_{ij}\Big]
\Big[ \left(g+\Pi_{ii}\right) f(z_{ij}) +\frac{f'(z_{ij})}{f(z_{ij})}\Pi_{ji}\Big] - \frac{\beta^2}{2} \sum_{i,j}^n \Pi_{ij}\Pi_{ji}.
\end{aligned}
\ee
Here, we abbreviated
\be
z_{ij}=x_i -x_j,
\ee
and the constant parameter $\beta$ and the function $f$ are given as follows,
\be\label{f}
\begin{aligned}
\mbox{rational Calogero model} \qquad &  \beta=0, \quad f(z_{ij}) =\frac{1}{z_{ij}}=\frac{1}{x_i{-}x_j}, \\
\mbox{hyperbolic  Calogero--Sutherland model} \qquad &  \beta=1, \quad f(z_{ij}) =\frac{1}{\sinh(z_{ij})}=\frac{1}{\sinh(x_i{-}x_j)}, \\
\mbox{trigonometric  Calogero--Sutherland model} \qquad &  \beta= \im, \quad f(z_{ij}) =\frac{1}{\sin(z_{ij})}=\frac{1}{\sin(x_i{-}x_j)}.
\end{aligned}
\ee

Thus, the supercharges \p{Q2} and the Hamiltonian \p{H2} describe an $\cN=2$-extended supersymmetric Calogero--Sutherland  models of type $A_{n-1}\oplus A_1$.

It should be noted that when checking that the supercharges form the superalgebra \p{N2SP} it is not enough to know the brackets
between $\Pi_{ij}$ and the fermions $\xi_{ij}, \bxi_{ij}$. Instead, the explicit expressions for $\Pi_{ij}$ \p{Pi0} have to be
substitute in the \p{Q2}. This makes the calculations slightly more complicated as comparing to those ones discussed in \cite{KLPS}.

\subsection{$\cN{=}\,4$ supersymmetric  $A_{n-1}\oplus A_1$ Calogero--Sutherland  models}
Due to the absence of any guiding rules for construction of $\cN{=}\,4$ supercharges, the reasonable starting point is the straightforward generalization of the $\cN{=}\,2$ supercharges \p{Q2} to the $\cN{=}\,4$ supersymmetry reads
\be\label{Q4}
\begin{aligned}
 & {Q}{}^a = \sum_{i=1}^n p_i \xi^a_{ii}\ -\ \im \sum_{i \neq j}^n \Bigl[ f(z_{ij})\left(g + \Pi_{jj}\right) +\frac{f'(z_{ij})}{f(z_{ij}) }\Pi_{ij}\Bigr]\, \xi^a_{ji} \ ,& \\
& {\bQ}_b  =  \sum_{i=1}^n p_i \bxi_{ii\,b}\ -\ \ \im \sum_{i \neq j}^n \Bigl[ f(z_{ij})\left(g + \Pi_{jj}\right) +\frac{f'(z_{ij})}{f(z_{ij}) }\Pi_{ij}\Bigr]\, \bxi_{ji\,b} ,&
\quad a,b =1,2\; .
 \end{aligned}
\ee
Unfortunately, this guess is not correct
and the supercharges \p{Q4} do not form the $\cN=4$ superalgebra 
\be\label{NSP}
\left\{ Q^a , \bQ_b \right\} = - 2 \im\, \delta^a_b\, H \und
\left\{ Q^a, Q^b \right\}=\left\{ \bQ_a, \bQ_b \right\}=0 \ .
\ee
 in contrast with their $\cN{=}\,2$ cousins \p{Q2} . The possible modification of the supercharges looks as
follows
\be\label{Q4fin}
\begin{aligned}
\cQ{}^a = {Q}{}^a - \im\,\beta\, \sum_{i,j}^n \xi^a_{ij}\tPi_{ji}, \qquad \overline{\cQ}{}_{a} ={\bQ}_{a} + \im\, \beta\, \sum_{i,j}^n \bxi_{ij\,a}\tPi_{ji}.
\end{aligned}
\ee
The supercharges \p{Q4fin} form $\cN{=}\,4$ super Poincar\'{e} algebra \p{NSP} together with the Hamiltonian
\be\label{H4}\
\begin{aligned}
\cH \ = \frac{1}{2}\sum_{i=1}^n p_i^2\ +\ \frac{1}{2}\sum_{i \neq j}^n \Big[ \left(g+\Pi_{jj}\right) f(z_{ij}) +\frac{f'(z_{ij})}{f(z_{ij})}\Pi_{ij}\Big]
\Big[ \left(g+\Pi_{ii}\right) f(z_{ij}) +\frac{f'(z_{ij})}{f(z_{ij})}\Pi_{ji}\Big] - \frac{\beta^2}{2} \sum_{i,j}^n \Pi_{ij}\Pi_{ji}.
\end{aligned}
\ee
which has the same form as $\cN{=}\,2$ Hamiltonian \p{H2}.

It should be mentioned that in the $\cN=2$ case the additional, $\beta$-dependent  terms in the supercharges \p{Q4fin} are automatically nullified 
in virtue of the structure of $\tPi_{ij}$ \p{Pi0}
\be
\begin{aligned}
\sum_{i,j}^n \xi^a_{ij}\tPi_{ji} = \sum_{i,j}^n \bxi^{ij\,a }\tPi_{ji} =0, \mbox{ if }  a,b =1,
\end{aligned}
\ee
and thus, the supercharges \p{Q4fin}
reduced to the supercharges \p{Q2} in the limit $a,b=1$.

Finally, all we said above is valid only for the functions $f$ from the list \p{f}.

\subsection{$\cN{=}\,4$ supersymmetric  $B_n, C_n$ and $D_n$ Calogero--Sutherland  models}
It is strange but for the $B$, $C$ and $D$-type models  the $\cN{=}\,4$ supercharges take the same form as $\cN{=}\,2$ ones.
Indeed, one may check that the following supercharges  (including $\tPi_{ij}$),
\be\label{BCDQ}
\begin{aligned}
\cQ{}^a & \ = \sum_{i=1}^n p_i \xi_{ii}^a -\im\, \sum_{i\neq j}^n \Big[ \left( g + \Pi_{jj}\right)  f(z_{ij})+ \frac{ f{}'(z_{ij})}{f(z_{ij})}\Pi_{ij}\Big] \xi^a_{ji}+\im 
\sum_{i\neq j}^n \Big[ \left( g + \Pi_{jj}\right)  f(y_{ij}) - \frac{ f{}'(y_{ij})}{f(y_{ij})}\tPi_{ij}\Big] \xi^a_{ji} \\
&\qquad + \im \sum_i^n \Big[ \left( g' + \Pi_{ii}\right)  f(y_{ii}) -\frac{ f{}'(y_{ii})}{f(y_{ii})}\tPi_{ii}\Big] \xi^a_{ii}, \\
\cbQ{}_a & \ = \sum_{i=1}^n p_i \bxi_{ii\,a} -\im\, \sum_{i\neq j}^n \Big[ \left( g + \Pi_{jj}\right)  f(z_{ij})+ \frac{ f{}'(z_{ij})}{f(z_{ij})}\Pi_{ij}\Big] \bxi_{ji\,a}-\im 
\sum_{i\neq j}^n \Big[ \left( g + \Pi_{jj}\right)  f(y_{ij}) - \frac{ f{}'(y_{ij})}{f(y_{ij})}\tPi_{ij}\Big] \bxi_{ji\,a} \\
&\qquad - \im \sum_i^n \Big[ \left( g' + \Pi_{ii}\right)  f(y_{ii}) -\frac{ f{}'(y_{ii})}{f(y_{ii})}\tPi_{ii}\Big] \bxi_{ii\,a}
\end{aligned}
\ee
form $\cN{=}\,4$ super Poincar\'{e} algebra \p{NSP} together with the Hamiltonian
\be\label{BCDH}
\begin{aligned}
\cH & = \ \frac{1}{2}\sum_{i=1}^n p_i^2\ +\ \frac{1}{2}\sum_{i \neq j}^n \Big[ \left(g+\Pi_{jj}\right) f(z_{ij}) +\frac{f'(z_{ij})}{f(z_{ij})}\Pi_{ij}\Big]
\Big[ \left(g+\Pi_{ii}\right) f(z_{ij}) +\frac{f'(z_{ij})}{f(z_{ij})}\Pi_{ji}\Big]\\
&\qquad +  \frac{1}{2}\sum_{i \neq j}^n \Big[ \left(g+\Pi_{jj}\right) f(y_{ij}) -\frac{f'(y_{ij})}{f(y_{ij})}\tPi_{ij}\Big]
\Big[ \left(g+\Pi_{ii}\right) f(y_{ij}) -\frac{f'(y_{ij})}{f(y_{ij})}\tPi_{ji}\Big] \\
&\qquad + \frac{1}{2}\sum_{i}^n \Big[ \left(g'+\Pi_{ii}\right) f(y_{ii}) -\frac{f'(y_{ii})}{f(y_{ii})}\tPi_{ii}\Big]
\Big[ \left(g'+\Pi_{ii}\right) f(y_{ii}) -\frac{f'(y_{ii})}{f(y_{ii})}\tPi_{ii}\Big]\\
&\qquad -\frac{\beta^2}{2} \sum_{i,j}^n \left( \Pi_{ij} \Pi_{ji} +  \tPi_{ij} \tPi_{ji}\right).
\end{aligned}
\ee
Here,
\be
y_{ij} = x_i+x_j ,
\ee
and the function $f$ is the same as in \p{f}. 

The  bosonic sector of the Hamiltonian \p{BCDH} reads
\be\label{bosBCDH}
\begin{aligned}
\cH_{bos}  = \frac{1}{2}\sum_{i=1}^n p_i^2\ +\frac{g^2}{2} \sum_{i\neq j}^n \left( f^2(z_{ij})+f^2(y_{ij})\right)+
\frac{g'{}^2}{2}\sum_{i}^n f^2(y_{ii}).
\end{aligned}
\ee
Due to the presence of only two coupling constants, $g$ and $g'$, we may describe $B$, $C$ and $D$-type models in the rational case and
$C$ and $D$ (but not $B$)-type models in the hyperbolic/trigonometric case.

Finally, let us noted that in the $\cN{=}\,2$ supersymmetric case the last term in the Hamiltonian \p{BCDH} nullified automatically due to structure of the 
$\Pi_{ij}$ and $\tPi_{ij}$ \p{Pi0} 
\be
\begin{aligned}
\sum_{i,j}^n \left( \Pi_{ij} \Pi_{ji} +  \tPi_{ij} \tPi_{ji}\right)=0 \qquad \mbox{ if } \qquad a,b=1.
\end{aligned}
\ee

\subsection{Towards higher supersymmetries}
It is interesting to note that the supercharges \p{Q4fin} obey the relations
\be\label{hN}
\begin{aligned}
\left\{ \cQ{}^a , \cQ{}^b \right\}=0 \quad \und \quad \left\{ \overline{\cQ}{}_{ a},\overline{\cQ}{}_{b}\right\}=0
\end{aligned}
\ee
for arbitrary range of the indices $a,b$ running from one to $\cN/2$. However, the anti-commutators between these supercharges have more complicated structure
\be\label{hN2}
\begin{aligned}
\left\{ \cQ{}_{ a},\overline{\cQ}{}_{b}\right\}= -2 \, \im \, \left( \delta^a_b \, \cH - \beta^2 \cV{}^a_b\right),
\end{aligned}
\ee
where
\be\label{V}
\begin{aligned}
\cV^a_b = \sum_{i,j}^n \left[ \Pi_{ij} W^a_{ji\,b}+\tPi_{ij} {\widetilde W}{}^a_{ji\,b}\right],
\end{aligned}
\ee
and
\be\label{W}
\begin{aligned}
 W^a_{ij\,b} = \sum_k^n \left[ \xi^a_{ik} \bxi_{kj\,b} + \bxi_{ik\,b}\xi^a_{kj}\right] \und
 {\widetilde W}{}^a_{ij\,b} = \sum_k^n \left[ \xi^a_{ik} \bxi_{kj\,b} - \bxi_{ik\,b}\xi^a_{kj}\right]\ .
 \end{aligned}
 \ee
 Thus, the algebra becomes nonlinear one. Note, that for the $\cN=2$ supersymmetry the unique term $\cV{}^1_1=0$, while for the
 $\cN=4$ supersymmetry we have $\cV{}^1_1=\cV{}^2_2$ and $\cV{}^1_2=\cV{}^2_1=0$. Thus, in the $\cN=4$ case these additional terms just
 modified the Hamiltonian.
 
 Finally, one should note that the purely fermionic objects $\cV{}^a_b$ \p{V} are, essentially, constants, because
 \be
 \left\{ \cH, \cV{}^a_b \right\} =0.
 \ee 

\setcounter{equation}{0}
\section{Euler--Calogero--Moser models}
\subsection{Basic ingredients}
The construction of the supersymmetric extension \cite{KLS2} of the Euler--Calogero--Moser systems \cite{sCal} is a more economical as comparing
to the supersymmetric Calogero--Sutherland systems we considered in the previous Sections. Indeed, to construct the corresponding $\cN$ supercharges
one needs to introduce ``only'' $\cN\,\frac{1}{2}n (n+1) $ fermions  $\rho^a_{ij}, \brho_{ij\,a}$ symmetric over indices ${i,j}$ 
$\rho^a_{ij}=\rho^a_{ji}, \, \brho_{a\, ij}=\brho_{a\,ji}$ and obeying the brackets
\be
\begin{aligned}
\left\{ \rho^a_{ij} , \brho_{km\, b}\right\} = -\frac{\im}{2}  \, \delta^a_b \left(  \delta_{im}\delta_{jk}+\delta_{ik}\delta_{jm}\right).
\end{aligned}
\ee
The  internal degrees of freedom of ECM models are encoded in the angular momenta $\ell_{ij} = -\ell_{ji}$ with the Poisson brackets forming the $so(n)$ algebra
\be\label{PB12}
\begin{aligned}
\left\{ \ell_{ij}, \ell_{km} \right\}= \frac{1}{2} \left(\delta_{ik} \ell_{jm}+\delta_{jm} \ell_{ik} -\delta_{jk} \ell_{im}-\delta_{im} \ell_{jk}\right).
\end{aligned}
\ee
Similarly to the construction of the supersymmetric Calogero--Sutherland systems, our anzats for the supercharges include the following
 fermionic bilinears\footnote{These bilinears now form an $su(n)$ algebra~\cite{KLS2}.}
\be\label{PiEOM}
\begin{aligned}
\Pi^\rho_{ij} & =  -\im \sum_{a=1}^{\cN/2} \;\sum_{k=1}^n \left( \rho^a_{ik}\brho_{kj\,a}-\rho^a_{jk}\brho_{ki\,a}\right), \quad
\left( \Pi^\rho_{ij}\right)^\dagger =\Pi^\rho_{ij},  \\
\tPi^\rho_{ij} & =  \sum_{a=1}^{\cN/2} \;\sum_{k=1}^{n } \left( \rho^a_{ik}\brho_{kj\,a}+\rho^a_{jk}\brho_{ki\,a }\right), \quad
\left( \tPi^\rho_{ij}\right)^\dagger =\tPi^\rho_{ij}.
\end{aligned}
\ee
Finally, one may check that the $\cN$ supercharges $Q^a, \bQ_b$
\be\label{Qecm}
\begin{aligned}
Q^a= \sum_{i=1}^n p_i \rho^a_{ii} - \sum_{i \neq j}^n \frac{\left( \ell_{ij}+\Pi^\rho_{ij}\right) \rho^a_{ji}}{x_i-x_j},\quad
\bQ_a= \sum_{i=1}^n p_i \brho_{ii\,a} - \sum_{i \neq j}^n \frac{\left( \ell_{ij}+\Pi^\rho_{ij}\right) \brho_{ji\, a}}{x_i-x_j}
\end{aligned}
\ee
form $\cN$-extended super Poincar\'{e} algebra \p{NSP} \cite{KLS2} together with the Hamiltonian
\be\label{Hecm}
\begin{aligned}
H= \frac{1}{2}\sum_{i=1}^n p_i^2 + \frac{1}{2} \sum_{i \neq j}^n \frac{\left( \ell_{ij}+\Pi^\rho_{ij}\right)^2 }{\left(x_i-x_j\right)^2}.
\end{aligned}
\ee

\subsection{$\cN{=}\,4$ supersymmetry}
The construction of the supersymmetric extension of the hyperbolic/trigonometric ECM model is very similar to the case
 of the Calogero model. Again, we succeeded in the construction of the $\cN{=}\,4$ supersymmetric extensions, only. The main idea of our
 construction is to maximally preserve the anzatz for the supercharges \p{Qecm}, i.e. we admit the appearance of the three-linear
 fermionic terms in the supercharges only through the bilinears \p{PiEOM}. Thus, our anzatz for the supercharges reads
\be\label{Qecmh}
\begin{aligned}
\cQ^a & =  \sum_{i=1}^n p_i \rho^a_{ii} - \sum_{i \neq j}^n \left( f(z_{ij})\, \ell_{ij} - \frac{f'(z_{ij})}{f(z_{ij})}\Pi^\rho_{ij}\, - \im\, \beta \tPi^\rho_{ij}\right)\rho^a_{ji}, \\
 {\overline \cQ}_{a} & = \sum_{i=1}^n p_i \brho_{ii\,a}   - \sum_{i \neq j}^n \left( f(z_{ij})\, \ell_{ij} - \frac{f'(z_{ij})}{f(z_{ij})}\Pi^\rho_{ij}\, + \im\, \beta \tPi^\rho_{ij}\right)\brho_{ji\,a},\ .
\end{aligned}
\ee
Here, the function $f(z_{ij})$ and the parameter $\alpha$  are defined in the list \p{f}.

In the rational case $f(z_{ij})=1/z_{ij}, \beta=0$ the supercharges \p{Qecmh} form $\cN$-extended  superalgebra \p{NSP} for the indices
$a,b= 1, \ldots \cN/2$. However, one may easily check that for the two other choices of $f(z_{ij}), \beta$ in \p{f}, the supercharges \p{Q4}  form only the $\cN{=}\,4$ superalgebra \p{NSP} together with the Hamiltonian
\be\label{Hecmh}
H=  \frac{1}{2}\sum_{i=1}^n p_i^2 + \frac{1}{2} \sum_{i \neq j}^n \left[ 
\left(f(z_{ij}) \ell_{ij}-\frac{f'(z_{ij})}{f(z_{ij})} \Pi^\rho_{ij}\right)^2 + \beta^2  \Pi^\rho_{ij}\Pi^\rho_{ji}\right] \ .
\ee

\section{Conclusions}
\vspace{0.5cm}
In this paper we constructed $\cN{=}\,4$ supersymmetric extensions of the Calogero--Sutherland models associated with the classical $A_n, B_n, C_n$ and $D_n$ series.  The guiding principle was the structure of the supercharges in which the fermion-cube terms are all built from fermionic bilinears \p{Pi0} spanning an $s(u(n)\oplus u(n))$ algebra. We also described $\cN{=}\,4$ supersymmetric extensions of the trigonometric/hyperbolic cousins of the Euler--Calogero--Moser system.

In contrast with the rational Calogero--Moser and/or Euler--Calogero--Moser system admitting an arbitrary number of supersymmetries \cite{KLS1,KLS2,KLPS}, their trigonometric/hyperbolic versions can so far be supersymmetrized up to the $\cN{=}\,4$ cases only. If we
try to construct the  additional supercharges, they will span a soft variant of the super Poincar\'{e} algebra with purely fermionic conserved R-charges
in the commutator of $\cQ^a$ with $\overline{\cQ}_b$~\p{hN2}. It will be interesting to understand the nature of these R-charges and their algebra in more detail.

For a further development, one of the key questions is the possible integrability or even super-integrability of the constructed systems. It seems there is no serious problem with the $L-A$ pairs, which mostly mimic the pairs from the bosonic case. However, the unusually large number of fermions complicates the situation. It should be noted that the Hamiltonians contain the fermions only
through the bilinears. Thus, the ``efficient'' number of the degrees of freedom seems to be smaller that the ``naive'' number of degrees of freedom. A qualitative example of such a situation is provided by the Euler--Calogero--Moser models~\cite{sCal},  where the internal degrees of freedom are encoded in currents spanning an $so(n)$ algebra.  
We will consider the integrability properties of the constructed supersymmetric systems elsewhere.

\section*{\bf Acknowledgements}
We thank Anton Sutulin for valuable discussions. 
This work was supported by the RFBF-DFG grant No 20-52-12003.

\end{document}